\documentclass{emulateapj}
\usepackage{graphicx}

\makeatletter
\newcommand*{\rom}[1]{\expandafter\@slowromancap\romannumeral #1@}
\newcommand{\pasa}{PASA}
\usepackage[backref,breaklinks,colorlinks,citecolor=black]{hyperref}
\newcommand{\OIII}{\hbox{[{\rm O}\kern 0.1em{\sc iii}]}}
\makeatother
\shorttitle{${\lowercase{z}}\sim2$: An Epoch of Disk Assembly}
\shortauthors{Simons et al.}

\usepackage{verbatim}

\begin{document}

\title{${\lowercase{z}}\sim2$: An Epoch of Disk Assembly}
\author{Raymond C. Simons\altaffilmark{1*}, Susan A. Kassin\altaffilmark{2}, Benjamin J. Weiner\altaffilmark{3}, Sandra M. Faber\altaffilmark{4},  Jonathan R. Trump\altaffilmark{5}, Timothy M. Heckman\altaffilmark{1}, David C. Koo\altaffilmark{4}, Camilla Pacifici\altaffilmark{2,6}, Joel R. Primack\altaffilmark{7}, Gregory F. Snyder\altaffilmark{2,8}, Alexander de la Vega\altaffilmark{1}}
\affil{$^1$Johns Hopkins University, Baltimore, 3400 North Charles St., MD, 21218, USA\\
	$^2$Space Telescope Science Institute, 3700 San Martin Dr., Baltimore, MD, 21218, USA\\
	$^3$Steward Observatory, 933 N. Cherry St., University of Arizona, Tucson, AZ 85721, USA\\
	$^4$UCO/Lick Observatory, Department of Astronomy and Astrophysics, University of California, Santa Cruz, CA 95064, USA\\
	$^5$Department of Physics, University of Connecticut, 2152 Hillside Rd Unit 3046, Storrs, CT 06269, USA\\
	$^6$NASA Postdoctoral Program Fellow\\
	$^7$Physics Department, University of California, Santa Cruz, CA 95064, USA\\
	$^8$Giacconi Fellow
	}

\begin{abstract}
We explore the evolution of the internal gas kinematics of star-forming galaxies from the peak of cosmic star-formation at $z\sim2$ to today. Measurements of galaxy rotation velocity $V_{rot}$, which quantify ordered motions, and gas velocity dispersion $\sigma_g$, which quantify disordered motions, are adopted from the DEEP2 and SIGMA surveys. This sample covers a continuous baseline in redshift from $z=2.5$ to $z=0.1$, spanning 10 Gyrs. At low redshift, nearly all sufficiently massive star-forming galaxies are rotationally supported ($V_{rot}>\sigma_g$). By $z=2$, the percentage of galaxies with rotational support has declined to 50$\%$ at low stellar mass ($10^{9}-10^{10}\,M_{\odot}$) and 70$\%$ at high stellar mass ($10^{10}-10^{11}M_{\odot}$). For $V_{rot}\,>\,3\,\sigma_g$, the percentage drops below 35$\%$ for all masses. From $z\,=\,2$ to now, galaxies exhibit remarkably smooth kinematic evolution on average. All galaxies tend towards rotational support with time, and it is reached earlier in higher mass systems.  This is mostly due to an average decline in $\sigma_g$ by a factor of 3 since a redshift of 2, which is independent of mass. Over the same time period, $V_{rot}$ increases by a factor of 1.5 for low mass systems, but does not evolve for high mass systems. These trends in $V_{rot}$ and $\sigma_g$ with time are at a fixed stellar mass and should not be interpreted as evolutionary tracks for galaxy populations. When galaxy populations are linked in time with abundance matching, not only does $\sigma_g$ decline with time as before, but $V_{rot}$ strongly increases with time for all galaxy masses. This enhances the evolution in $V_{rot}/\sigma_g$. These results indicate that $z\,=\,2$ is a period of disk assembly, during which the strong rotational support present in today's massive disk galaxies is only just beginning to emerge.
\end{abstract}

\keywords{galaxies: evolution - galaxies: formation -galaxies: fundamental parameters - galaxies: kinematics and dynamics}

\let\thefootnote\relax\footnote{$^*$rsimons@jhu.edu}

\section{Introduction}
The peak of cosmic star-formation at $z\sim2$ is expected to be a violent period for galaxies. Several active processes, including minor and major mergers, smooth accretion, feedback from star-formation, and violent disk instabilities, can disrupt or stall disk formation. Indeed, observations indicate that the internal kinematics of star-forming galaxies at this time have large amounts of disordered motions (as measured by the velocity dispersion of the gas $\sigma_g$) in addition to ordered rotation ($V_{rot}$) (e.g., \citealt{2006Natur.442..786G, 2006ApJ...645.1062F, 2007ApJ...669..929L, 2007ApJ...658...78W, 2008ApJ...687...59G, 2008ApJ...682..231S, 2009ApJ...697.2057L, 2009ApJ...706.1364F, 2011ApJ...742...11S, 2012Natur.487..338L, 2013ApJ...767..104N, 2013PASA...30...56G, 2015ApJ...799..209W, 2016ApJ...819...80P, 2016ApJ...830...14S, 2017ApJ...838...14M, 2017ApJ...839...57S}). Even still, the majority of massive star-forming galaxies, $\log M_*/M_{\odot}\,>\,10$, at this redshift are rotation-dominated ($V_{rot}\,>\,\sigma_g$), with at least 70$\%$ showing disk-like kinematic signatures \citep{2015ApJ...799..209W}.

However, these galaxies are unlike local disks. Locally, massive star-forming galaxies have much stronger rotational support ($V_{rot}/\sigma_g\sim10$) and low values of $\sigma_g$ ($\sim20$ km s$^{-1}$). At a redshift of 2, the quantity $V_{rot}/\sigma_g$ rarely exceeds a few, indicating that these galaxies mature significantly in the 10 Gyrs between $z\,=\,2$ and now. 

The evolution of $V_{rot}$ and $\sigma_g$ from $z\,=\,1.2$ to now was studied by \citet{2012ApJ...758..106K} using the DEEP2 survey \citep{2013ApJS..208....5N}. Over this period, they found that galaxy disks settle with time, increasing in $V_{rot}/\sigma_g$ by increasing in $V_{rot}$ and declining in $\sigma_g$. This evolution is a function of mass, with massive galaxies developing strong rotational support first (i.e., ``kinematic downsizing"; \citealt{2012ApJ...758..106K}).

Although significant progress has been made in understanding the kinematic state of star-forming galaxies at even higher redshifts, it is unclear how these lower redshift trends from \citet{2012ApJ...758..106K} relate to the large amounts of disordered motions found in galaxies at $z=2$. One might not expect a smooth extension to $z\,=\,2$, where the processes governing galaxy assembly are violent and inhospitable to disk formation.

To examine this, or any kinematic evolution, it is important to have a large and homogenous sample. In addition, the sample must have minimal biases, cover the same galaxy mass range at all redshifts, and ideally have kinematics measured using the same technique. To address these factors, we combine the kinematics samples from the DEEP2 survey (\citealt{2012ApJ...758..106K}), which spans $0.1\,<\,z\,<\,1.2$, and the SIGMA survey \citep{2016ApJ...830...14S}, which spans $1.3\,<\,z\,<\,2.5$. These samples have similar selections. They are both morphologically unbiased (i.e., they do not select only disk-like systems) and they trace the star-formation main sequence. Both samples contain enough galaxies to overcome the large intrinsic scatter in galaxy kinematics. Furthermore, measurements were made using the same fitting routine in both surveys. The combined sample covers a large redshift range, $0.1\,<\,z\,<\,2.5$, and spans a significant portion of the age of universe, from 2.6 to 12.4 Gyr. In this paper, we quantify the evolution of the following properties: $V_{rot}$, $\sigma_g$ and their contributions to the total dynamical support. Moreover, we examine how these properties depend on stellar mass.

The format of this paper is as follows. First, in \S 2, we summarize the data sets used and discuss the measurements of kinematics and stellar masses. In \S 3 and 4 we examine the evolution of galaxy kinematics and the fraction of rotationally supported systems in our sample, respectively, at a fixed stellar mass. Next, in \S 5, we link galaxy populations in time and examine their kinematic evolution. Finally, we summarize our conclusions in \S 6. We adopt a flat $\Lambda$CDM cosmology defined with (h, $\Omega_m$, $\Omega_{\Lambda}$) = (0.704, 0.272, 0.728).

\section{Data}
\begin{figure*}
\begin{centering}
\includegraphics[angle=0,scale=.85]{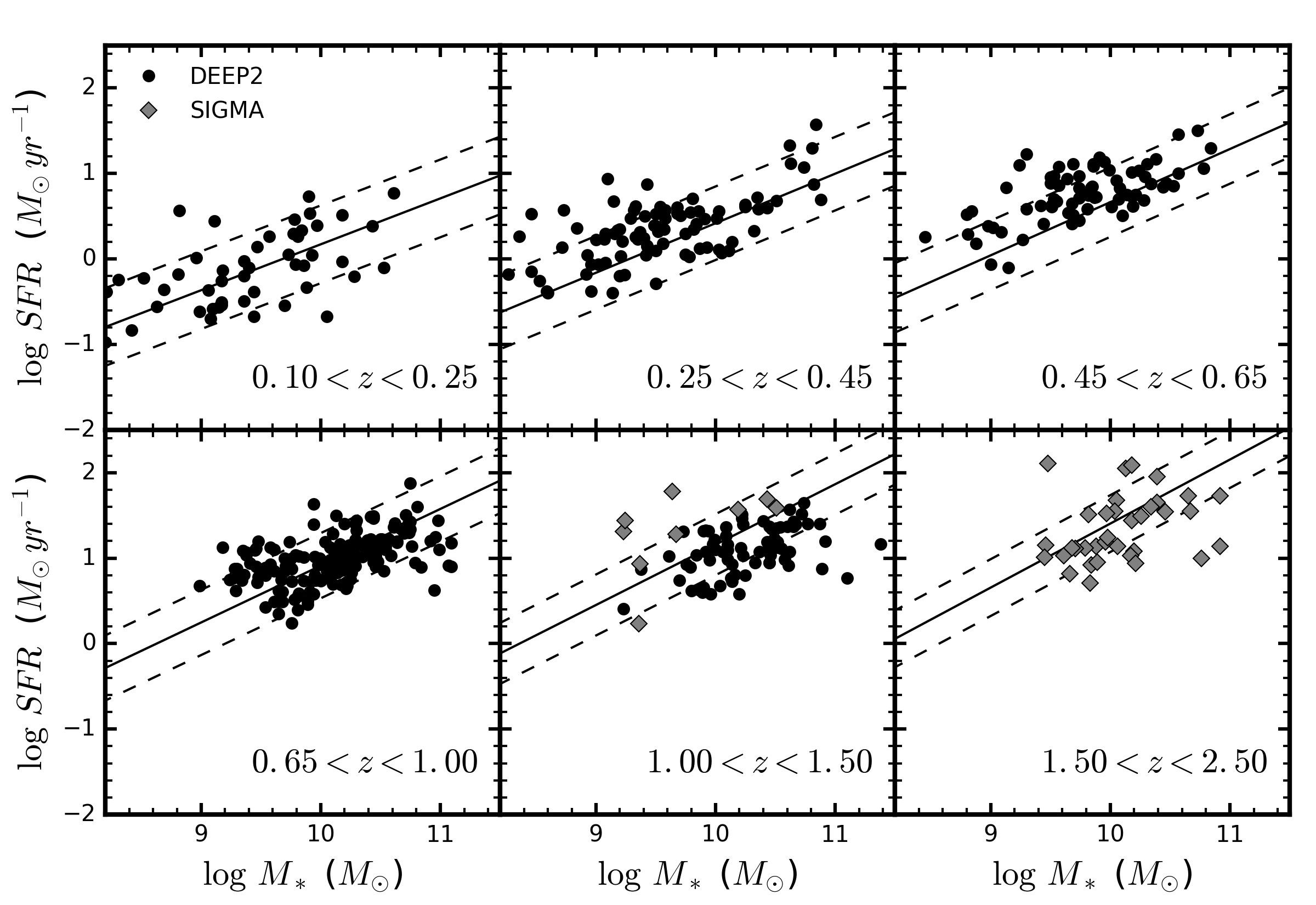}
\caption {Galaxies in the DEEP2 (black circles) and SIGMA (grey diamonds) kinematics samples lie along the star-formation main sequence at their respective redshifts. The panels show redshift bins in equal intervals of lookback time. The solid and dashed lines are the main sequence fit and rms scatter for each redshift bin, respectively, from \citealt{2014ApJS..214...15S} and renormalized to a Chabrier IMF.\vspace{0.1cm}}
\label{fig:sfr_m}
\end{centering}
\end{figure*}

Measurements of the internal kinematics of star-forming galaxies are adopted from \citet{2012ApJ...758..106K} (from the DEEP2 survey) and \citet{2016ApJ...830...14S} (from the SIGMA survey). Both samples are morphologically unbiased (i.e., not only disk-like systems are selected) and kinematics are measured from rest-optical emissions lines which trace the hot ionized $T\sim10^4$ K gas in galaxies. The galaxies used in this paper are shown in a stellar mass vs. star formation rate ($M_*-\,SFR$) diagram in Figure \ref{fig:sfr_m}. We compare with the star-formation main sequence at their respective redshifts (from \citealt{2014ApJS..214...15S}) and show that our sample is representative of typical star-forming galaxies over $0.1\,<\,z\,<\,2.5$. We note that the galaxies over the redshift range $0.3\,<\,z\,<\,0.7$ tend to lie slightly above the main sequence. However, the majority of the sample is within the 1$\sigma$ scatter of the relation at all redshifts. 

Below we describe the DEEP2 and SIGMA samples (\S 2.1 and 2.2, respectively), as well as the measurements of kinematics (\S 2.3), stellar masses and star-formation rates (\S 2.4).

\begin{figure*}
\includegraphics[angle=0,scale=1.15]{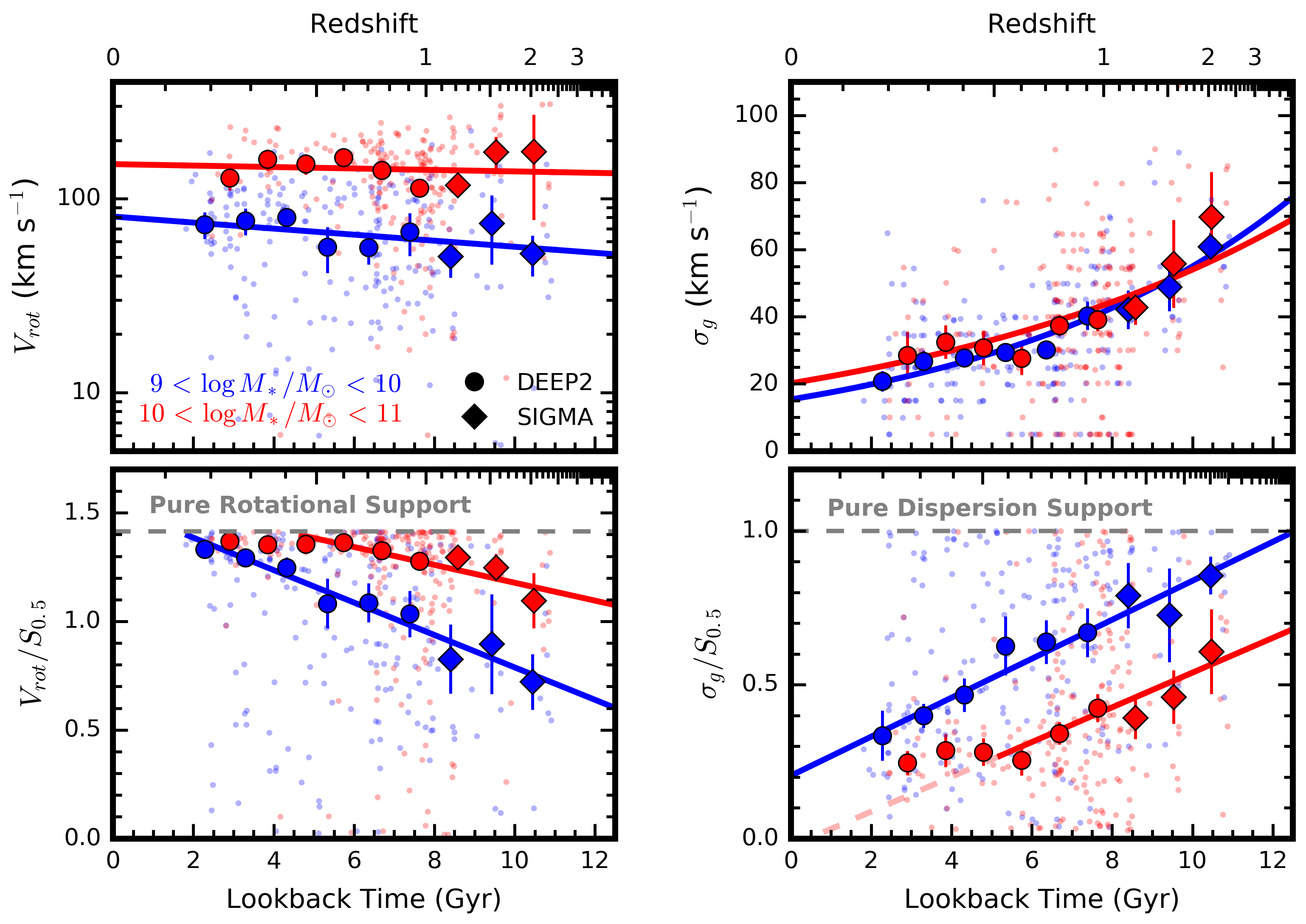}
\caption{Star-forming galaxies evolve in $V_{rot}$ and $\sigma_g$ with time. The small faint background points are measurements for individual galaxies. The large points and associated error bars show medians of the individual points in bins of lookback time and their standard error. Solid lines are the best-fit relations to the median points. Low mass and high mass galaxies are shown in blue and red, respectively. Top left: The average rotation velocity, $V_{rot}$, increases with time since $z\,=\,2.5$ for low mass galaxies, but shows no evolution for high mass galaxies. Top right: The gas velocity dispersion $\sigma_g$, which traces disordered motions, decreases precipitously from $z\,=\,2.5$ to today for both low and high mass galaxies. In the bottom panels, $V_{rot}$ and $\sigma_g$ are normalized by $S_{0.5}$, which traces the total dynamical support of galaxies. This allows us to examine the fraction of total dynamical support that $V_{rot}$ and $\sigma_g$ provide to galaxies. At $z\,\sim\,2$, low mass galaxies have a significant fraction of their total support in disordered motions. With time, all galaxies on average increase in rotational support (i.e., increase in $V_{rot}/S_{0.5}$) and decrease in dispersion support (i.e., decrease in $\sigma_g/S_{0.5}$). This happens earlier for higher mass galaxies.\vspace{0.1cm}}
\label{fig:sig_v_S_z}
\end{figure*}

\subsection{DEEP2 sample}
 
The intermediate redshift sample ($0.1\,<\,z\,<\,1.2$) used in this paper is from \citet{2007ApJ...660L..35K} and \citet{2012ApJ...758..106K} (hereafter K12). The sample is briefly discussed here and the reader is referred to K12 for further details. The galaxies in the K12 sample are located in the Extended Groth Strip. Spectra, from which the internal kinematics are measured, are from the DEEP2 survey \citep{2013ApJS..208....5N} and were taken with the DEIMOS instrument \citep{2003SPIE.4841.1657F} on the Keck-II telescope. The slits used were 1$\arcsec$ wide and the spectral resolution is $R\,\sim\,$5000 ($\sigma_{instr} = 26$ km s$^{-1}$). { The nominal on-source exposure times were 1 hour.} The spectra were observed in natural optical seeing conditions, which varied between 0.5$\arcsec$ and 1.2$\arcsec$. Galaxies are selected to have bright enough emission lines to measure kinematics ($\gtrsim10^{-17}$ erg s$^{-1}$ cm$^{-2}$). Additionally, galaxies are required to have $V$- and $I$-band imaging from the Hubble Space Telescope/Advanced Camera for Surveys (HST/ACS) instrument. Inclinations were measured from the Hubble imaging using the {\tt{SExtractor}} program \citep{1996A&AS..117..393B} in \citet{2008ApJ...672..177L}. Galaxies are selected to have inclinations between $30^{\circ}\,<\,i\,<\,70^{\circ}$, avoiding dust effects in edge-on systems and highly uncertain inclination corrections in face-on systems. Furthermore, galaxies are required to have slits aligned within 40$^{\circ}$ of the photometric major axes \citep{2006ApJ...653.1027W, 2010ApJ...710..279C}. Following \citet{2015MNRAS.452..986S}, we apply a conservative selection on the rest-optical HST half-light diameter ($D_{50}\,>\,0.8$ times the FWHM of the seeing). This size selection removes 50 galaxies. The sample used in this paper contains 462 galaxies.

\subsection{SIGMA sample}
The high redshift ($1.3\,<\,z\,<\,2.5$) sample used in this paper is from the SIGMA kinematics survey (\citealt{2016ApJ...830...14S}; hereafter S16). The sample is briefly discussed here and we refer to S16 for further details. The galaxies in SIGMA are located in the UDS, GOODS-S and GOODS-N fields. Spectra were taken with the MOSFIRE spectrograph \citep{2010SPIE.7735E..1EM, 2012SPIE.8446E..0JM} on the Keck-I telescope as a part of the TKRS-2 survey \citep{2015AJ....150..153W} and were also published in \citet{2013ApJ...763L...6T} and \citet{2014ApJ...795..145B}. The slits were 0.7$\arcsec$ wide and the spectral resolution is $R\,\sim\,$3630 ($\sigma_{inst} = 35\,$km s$^{-1}$). { The on-source exposure times ranged between 1.5 - 2 hours and the near-infrared seeing varied between 0.45 - 0.85$\arcsec$.} HST/Wide Field Camera 3 (WFC3) imaging is available for all of the galaxies in SIGMA through the CANDELS survey \citep{2011ApJS..197...35G, 2011ApJS..197...36K}. Axis ratios were measured from the HST/WFC3 $H$-band image using the {\tt GALFIT} software \citep{2010AJ....139.2097P} by \citet{2012ApJS..203...24V} and are used to derive inclinations. As in the K12 sample, all galaxies are required to have at least one slit aligned with the photometric position angle and intrinsic emission sizes which are large enough to resolve kinematics ($D_{50}\,>\,0.8$ times the FWHM of the seeing; \citealt{2016ApJ...830...14S}). We remove four galaxies from the original sample for which inclinations corrections were not applied. The sample used in this paper contains 45 galaxies and a catalog of their physical properties and kinematic measurements is available in S16.

\subsection{Kinematic Measurements}

Kinematics (rotation velocity $V_{rot}$, and gas velocity dispersion $\sigma_g$) were measured from strong nebular emission lines  (H$\alpha$ and \OIII$\lambda$5007) with the {\tt ROTCURVE} program \citep{2006ApJ...653.1027W} in \citet{2007ApJ...660L..35K} and S16 for the DEEP2 and SIGMA surveys, respectively. The measurement technique is briefly discussed here and the reader is referred to \citet{2006ApJ...653.1027W} for details.

Seeing blurs rotation and artificially elevates the velocity dispersion in the center of the galaxy (due to the rise in the rotation curve) and needs be taken into account in the kinematic modeling. {\tt ROTCURVE} models the velocity profile of the emission line, taking into account seeing, with two components: an arctangent rotation curve and a constant dispersion term. The rotation velocity uncorrected for inclination ($V_{rot}\times\sin i$) is measured at the flat portion of the rotation curve. { Recent evidence has suggested that the rotation curves of high redshift star-forming galaxies fall, instead of flatten, beyond 1.5 effective radii \citep{2017arXiv170305491L, 2017Natur.543..397G}. Our high redshift data do not extend far enough to distinguish between these two cases. However, there is only a small difference in the maximum velocity derived from a fit to either model. Adopting a value of $V_{rot}$ which is measured at a smaller radius, e.g., at 1.5 R$_{e}$, would lower the rotation velocities by $\sim0.1$ dex across our full sample. For consistency across redshift, we adopt $V_{rot}$ as measured at the flat part of the modeled rotation curve in all of our galaxies.}

Typical uncertainties on measurements of $V_{rot}\times\sin i$ are approximately 10 km s$^{-1}$ for DEEP2 and 30 km s$^{-1}$ for SIGMA. Uncertainties on $\sigma_g$ are approximately 15 km s$^{-1}$ for DEEP2 and 25 km s$^{-1}$ for SIGMA. Inclination corrections are applied to the {\tt ROTCURVE} values using the rest $\sim V$-band axis ratio measured from the Hubble images (ACS $V$- and $I$-bands in K12 for $0.1\,<\,z\,<\,0.6$ and $0.6\,<\,z\,<\,1.2$, respectively, and WFC3 $H$-band in S16 for $1.3\,<\,z\,<\,2.5$). 

The gas velocity dispersion $\sigma_g$ integrates small scale velocity gradients below the seeing limit \citep{2010ApJ...710..279C, 2014ApJ...790...89K}. It is mostly due to non-ordered motions among HII regions, but also includes smaller contributions from thermal broadening ($\sim$10 km s$^{-1}$ for H gas at 10$^4$ K) and the internal turbulence in HII regions ($\sim$20 km s$^{-1}$ in local HII regions, \citealt{1990ARA&A..28..525S}). As in K12, we refer to $\sigma_g$  as tracing ``disordered motions''. 

The quantity $V_{rot}$ measures the ordered motions of a galaxy while the quantity $\sigma_g$ measures its disordered motions. We follow \citet{2006ApJ...653.1027W} and \citet{2007ApJ...660L..35K} by combining both forms of dynamical support into the quantity $S_{0.5}=\sqrt{0.5\,V_{rot}^2+\sigma_g^2}$. This term serves as a kinematic tracer of the total mass in an isothermal potential \citep{2006ApJ...653.1027W}. 

\begin{figure}
\begin{centering}
\includegraphics[angle=0,scale=1.1]{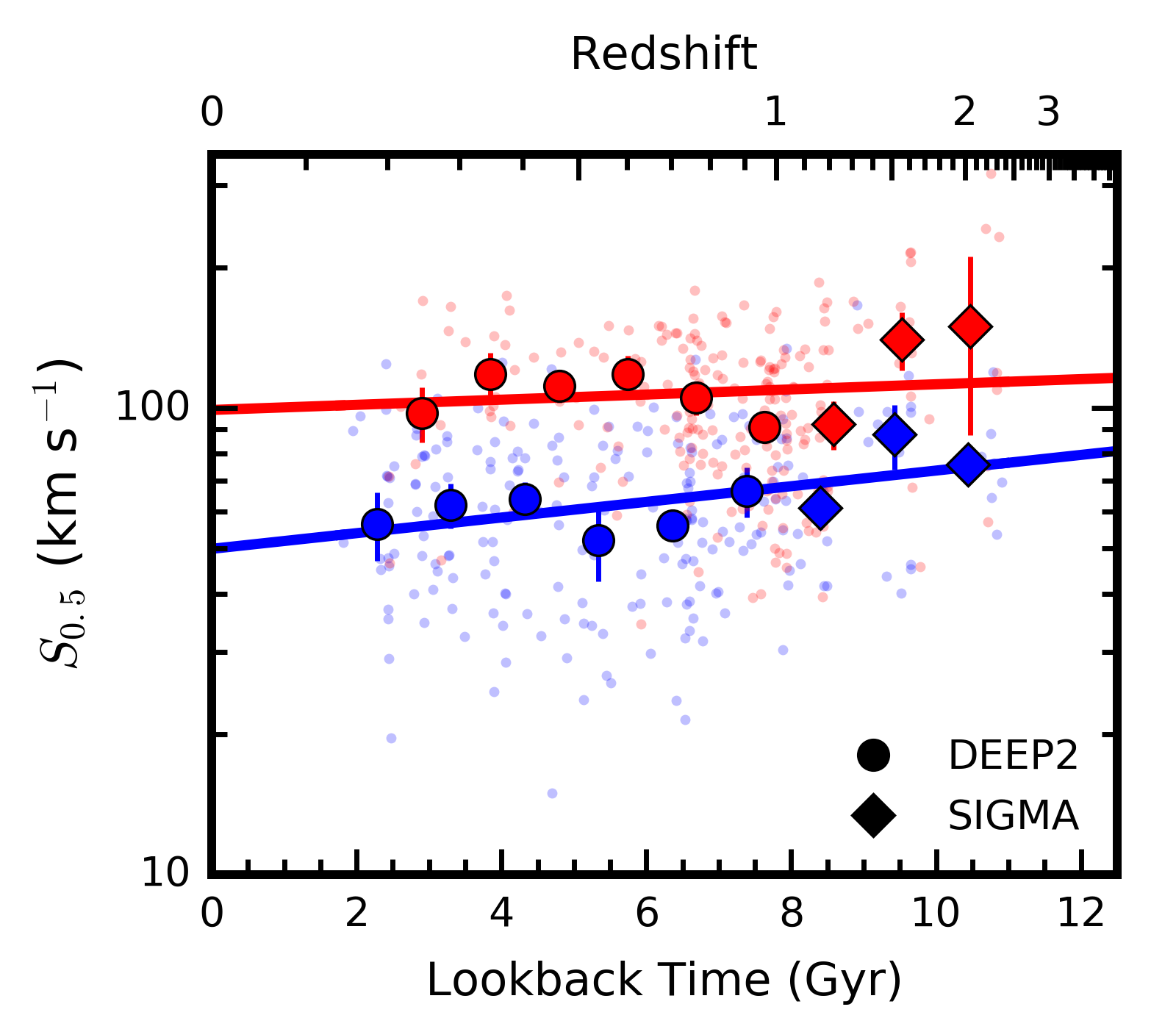}
\caption {The evolution of $S_{0.5}$, a tracer of the galaxy potential well depth, is shown. This quantity mildly declines since $z=2.5$ in galaxies with low stellar mass (blue), which is largely driven by the offset of the two highest redshift points. $S_{0.5}$ has not changed significantly in high mass galaxies (red). The points and lines are the same as in Figure \ref{fig:sig_v_S_z}.\vspace{0.1cm}}
\label{fig:S_time}
\end{centering}
\end{figure}

\subsection{Stellar Mass and SFR Measurements}

Stellar masses ($M_*$) were measured for the galaxies in the K12 sample using absolute B-band magnitudes and rest B$-$V colors \citep{2001ApJ...550..212B, 2005ApJ...625...23B} with empirical corrections from spectral energy distribution (SED) fitting \citep{2006ApJ...651..120B}, as in \citet{2007ApJ...660L..51L}. A \citet{2003PASP..115..763C} initial mass function was adopted and the uncertainties in stellar mass are approximately 0.2 dex. Star-formation rates (SFR) were derived in \citet{2007ApJ...660L..43N}. For galaxies with 24$\micron$ detection, the obscured SFR was measured from the IR luminosity using the SED templates of \citet{2001ApJ...556..562C} and is added to the unobscured component of the SFR, derived from the DEEP2 emission line luminosities \citep{2007ApJ...660L..39W}. Otherwise, the SFR was measured from the extinction corrected emission line luminosities following the calibration in \citet{1998ARA&A..36..189K}.

Stellar masses for the galaxies in SIGMA were measured from SED fitting to the UV-optical-NIR data available in the CANDELS fields, as described in \citet{2014ApJ...795..145B}. The fit was performed with {\tt FAST} \citep{2009ApJ...700..221K} assuming a \citet{2003PASP..115..763C} initial mass function, \citet{2003MNRAS.344.1000B} stellar population synthesis models and a Calzetti extinction law \citep{2000ApJ...533..682C}. The uncertainties in stellar mass are approximately 0.3 dex \citep{2015ApJ...808..101M}. For galaxies with detections in the mid-to-far IR, the SFR was calculated using both the obscured (IR) and unobscured (UV) components (\citealt{1998ARA&A..36..189K}). Otherwise, SFRs are derived from the extinction corrected UV using the dust parameters of the best fit SED model. 

\section{Kinematic Evolution At Fixed Stellar Mass}

We use measurements of $V_{rot}$, $\sigma_g$, and $S_{0.5}$ from the K12 and S16 samples to explore the kinematic evolution of star-forming galaxies from $z\sim2.5$ to $z\sim0.1$. Bins in equal intervals of lookback time ($t_{L}$) are adopted, each spanning 1 Gyr. Our conclusions are insensitive to the size of these intervals. We first examine evolution at fixed stellar mass for two stellar mass ranges: $9\,<\,\log M_*/M_{\odot}\,<\,10$ (hereafter referred to as ``low mass") and $10\,<\,\log M_*/M_{\odot}\,<\,11$ (hereafter referred to as ``high mass"). The $z\,\sim\,2.2$ ($t_{L} \approx 10$ Gyr) bin in S16 does not include galaxies below $\log\,M_{*}/M_{\odot}\,=\,9.4$ and instead encompasses $9.4\,<\,\log\,M_{*}/M_{\odot}\,<\,10$. The high stellar mass bin at $z\,\sim\,0.3$ is not included as it contains fewer than five galaxies. The removal or inclusion of these bins does not affect our results. We briefly summarize the results of this section below and refer to the subsections for details.

We find a large amount of scatter in the individual measurements of $V_{rot}$, $\sigma_g$ and $S_{0.5}$ at a given mass but smooth {\emph{average}} trends with mass and time, highlighting the need for a large sample. On average, the rotation velocities of low mass galaxies rise by a factor of 1.5 from $z\,\sim\,2.5$ to the present day. High mass galaxies show little to no evolution in $V_{rot}$ over the same period. The gas velocity dispersion smoothly declines from $z\,\sim\,2.5$ to today at a remarkably similar rate in both mass bins. We consider contributions from $V_{rot}$ and $\sigma_g$ to the total dynamical support, as quantified by $S_{0.5}$.  While all galaxies tend towards rotational support with time, it is obtained earlier for galaxies with higher stellar masses. This phenomenon, dubbed ``kinematic downsizing'', was first shown by K12 to $z=1.2$ and we extend it to $z\,=\,2.5$. This result follows from the facts that $\sigma_g$ does not depend significantly on mass and low mass galaxies sit in shallower potential wells than high mass galaxies. 

Low mass galaxies are delayed in their kinematic development, taking an additional 4 - 5 Gyr to reach the same level of rotational support as high mass galaxies. From $z\,=\,2$ to $z\,=\,1$, low mass galaxies are strongly supported by dispersion ($\sigma_g/S_{0.5}\,>\,0.7$). By $z\,=\,0.7$, high mass galaxies have developed strong rotational support ($\sigma_{g}/S_{0.5}\,<\,0.3$), while low mass galaxies are only reaching a similar level of rotational support today.

It is important to note that galaxies grow in mass over time and so these trends at fixed stellar mass should not be interpreted as tracks for individual galaxy populations.

\subsection{Increase in $V_{rot}$ With Time for Low Mass Galaxies}

The evolution of the rotation velocity, $V_{rot}$, is shown in the top left panel of Figure \ref{fig:sig_v_S_z}. The galaxies in our sample span a large range in $V_{rot}$ at fixed epoch. This scatter is mostly intrinsic but is due in part to our relatively wide bins in stellar mass (1 dex) and to a lesser extent from measurements uncertainties. 

At all times, the average rotation velocity is higher for galaxies with higher stellar mass. We find a mild increase of the average $V_{rot}$ with time in galaxies with low stellar mass, rising from 50 km s$^{-1}$ at $z\,=\,2$ to 70 km s $^{-1}$ at $z\,=\,0.2$. High mass galaxies show little evolution in $V_{rot}$ over the same time period. We estimate the standard error on the median $V_{rot}$ in each bin through bootstrap resampling and perform an uncertainty weighted least squares fit to the median points with a line:

\begin{equation}\label{eq:1}
\log \left(\frac{V_{rot}}{\textrm{km s$^{-1}$}}\right) = a \left(\frac{t_{L}}{\text{Gyr}}\right)+b
\end{equation}
The best-fit values of the slope and intercept are $a\,=-0.02\pm 0.01$ (i.e., slow rise) and $b\,=1.91\pm 0.07$ for our low stellar mass bin, and $a\,=-0.004\pm 0.014$ (i.e., no evolution) and $b\,=2.18\,\pm\, 0.09$ for our high stellar mass bin. The intercept value of each mass bin, i.e., $V_{rot}$ at $z\,=\,0$, is consistent with the local stellar mass Tully-Fisher relation (90 and 175 km s$^{-1}$ at $\log M_*/M_{\odot}= 9.5$ and $10.5$, respectively; \citealt{2011MNRAS.417.2347R}).

\subsection{Smooth Decay of $\sigma_g$ With Time for All Galaxies}

The evolution of the gas velocity dispersion, $\sigma_g$, is shown in the top right panel of Figure \ref{fig:sig_v_S_z}. As with $V_{rot}$, galaxies in our sample span a wide range in $\sigma_g$ at all epochs, largely due to measurement uncertainties. 

The galaxies at the highest redshifts have values of $\sigma_g$ which are roughly 3 times larger than the galaxies at the lowest redshifts (see also \citealt{2015ApJ...799..209W}). The quantity $\sigma_g$ decreases from 70 km s$^{-1}$ to 20 km s$^{-1}$ over $0.2\,<\,z\,<\,2.0$. As first shown in K12, $\sigma_g$ roughly doubles over $0.1\,<z\,<1.2$, which spans approximately 8 Gyrs. Although the time from $z\,=\,1$ to $z\,=\,2$ is much shorter, only 3 Gyrs, we find significant evolution in $\sigma_g$ during this period, with the median increasing from 40 km s$^{-1}$ to 70 km s$^{-1}$. We perform a fit to the median $t_{L}-\sigma_g$ relation for our two mass bins with:

\begin{equation}\label{eq:2}
\log \left(\frac{\sigma_g}{\textrm{km s$^{-1}$}}\right) = a\left(\frac{t_{L}}{\textrm{Gyr}}\right)+b
\end{equation}
The best fit values are $a\,=\,0.055\,\pm\,0.010$ and $b\,=\,1.19\,\pm\,0.04$ for the low mass bin and $a\,=\,0.043\,\pm\,0.010$ and $b\,=\,1.31\,\pm\,0.07$ for the high mass bin. There is no significant difference between the evolution of $\sigma_g$ in the low mass and high mass bins. {\emph{This result implies a mass-independent half-life timescale for $\sigma_g$, the time over which it declines by a factor of 2, of approximately 6 Gyrs.}} The intercept values, i.e., $\sigma_g$ at $z\,=\,0$, are consistent with measurements of the ionized gas velocity dispersion in local star-forming disk galaxies ($\sim15-25$ km s$^{-1}$, e.g., \citealt{2008MNRAS.390..466E}).

\begin{figure*}[ht]
\begin{centering}
\includegraphics[angle=0,scale=.42]{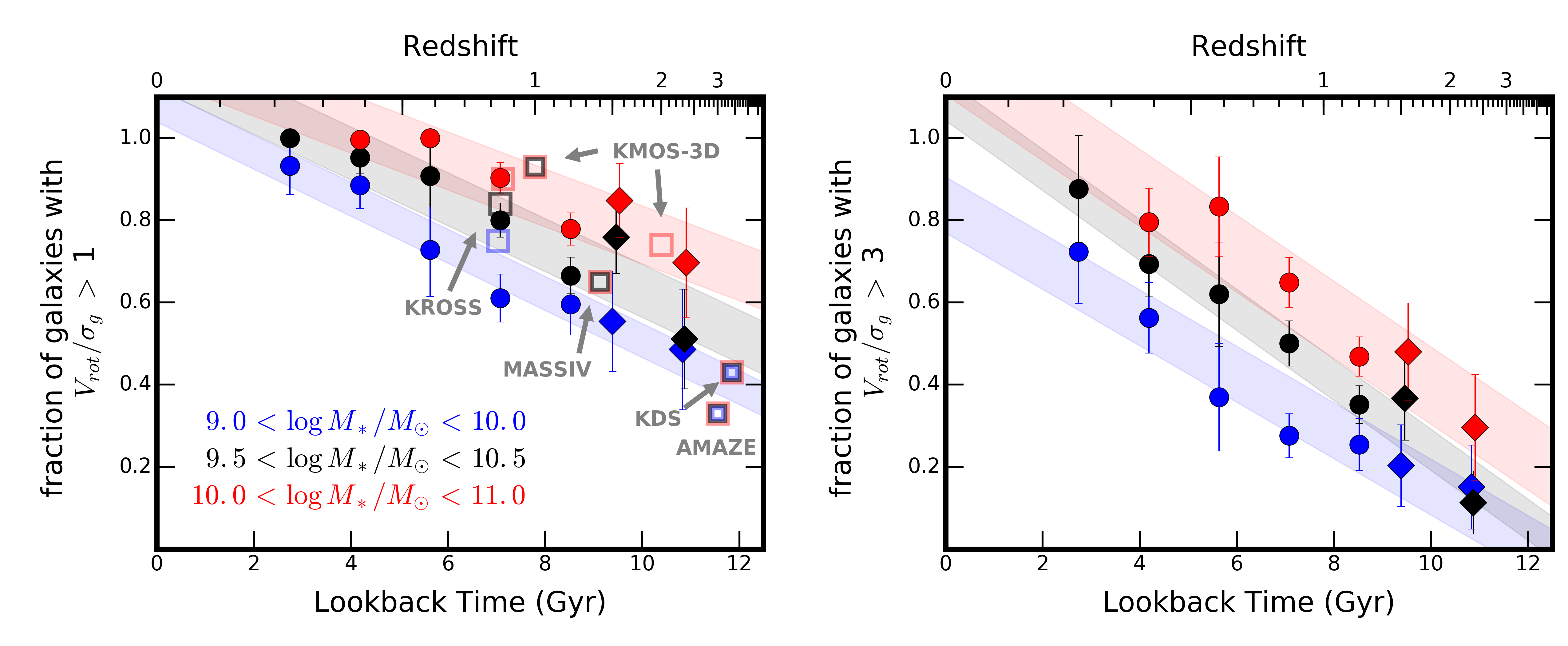}
\caption {The fraction of star-forming galaxies with disk-like kinematics declines with increasing redshift and decreasing mass. In the left panel, the fractions of galaxies with $V_{rot}/\sigma_g\,>\,1$ are shown as a function of lookback time and redshift (bottom and top x-axes, respectively) and galaxy stellar mass. The solid points show measurements from DEEP2 (circles) and SIGMA (diamonds) and the error bars are measured via bootstrap resampling. The shaded swaths show the uncertainty on the intercept in a linear fit to the points. By $z\,=\,2$, only $\sim50-70\%$ of all galaxies meet the very lenient criteria of $V_{rot}/\sigma_g\,>\,1$. In the right panel, the fractions of galaxies with $V_{rot}/\sigma_g\,>3$ are shown. By $z\,=\,2$, less than $40\%$ of all galaxies meet this criterion. Measurements from kinematics surveys in the literature are shown as open squares and are color-coded by their mass ranges with the same stellar mass bins used for our data. Samples which span multiple bins in mass are shown with two or more colors. We find good agreement among all surveys once stellar mass is accounted for.\vspace{0.1cm}}
\label{fig:mass_settling}
\end{centering}
\end{figure*}

\subsection{Evolution of Dynamical Support, $S_{0.5}$, With Time}

In Figure \ref{fig:S_time}, we examine the evolution of $S_{0.5}$, which traces the total dynamical support of galaxies. We fit a line to the median $S_{0.5}$ vs. lookback time:

\begin{equation}\label{eq:3}
\log \left(\frac{S_{0.5}}{\textrm{km s$^{-1}$}}\right) = a\left(\frac{t_{L}}{\textrm{Gyr}}\right)+b
\end{equation}

The best-fit slope and intercept values are $a=0.017\,\pm\,0.008$ and $b=1.70\,\pm\,0.06$ for the low mass bin and $a=0.01\,\pm\,0.01$ and $b=2.00\,\pm\,0.09$ for the high mass bin. The quantity $S_{0.5}$ has marginally declined in low mass galaxies since $z\,=\,2.5$, i.e., they now sit in shallower potential wells, but we measure no significant evolution in high mass galaxies. 

While the fit to the low mass galaxies indicates a smooth evolution, we caution that the trend is largely driven by the high redshift measurements. The median points below $z\,=\,1.5$ are consistent with a flat slope. The points above $z\,=1.5$ are all above this flat line, consistent with a zero-point shift in the modified Tully-Fisher relation at high redshift (\citealt{2009ApJ...697..115C, 2016ApJ...819...80P, 2017ApJ...839...57S, 2017arXiv170304321U}).

 \subsection{Fraction of Total Dynamical Support Given By $V_{rot}$ and $\sigma_g$ Over Time}
 
We normalize $V_{rot}$ and $\sigma_g$ by $S_{0.5}$ and measure the relative contributions of ordered and disordered motions, respectively, to the total dynamical support. A thin disk which is supported by rotation will tend toward $V_{rot}/S_{0.5}\,=\,\sqrt{2}$, and a system which is supported by dispersion will tend toward $\sigma_g/S_{0.5}\,=\,1$. 

The evolution of $V_{rot}/S_{0.5}$ is shown in the bottom left panel of Figure \ref{fig:sig_v_S_z}. K12 showed that $V_{rot}/S_{0.5}$ rises from $z\,=\,1.2$ to now and is a strong function of mass. We find a smooth extension of these results to $z\,=\,2.5$. 

At all redshifts, high mass galaxies have higher values of $V_{rot}/S_{0.5}$, i.e., they are more rotationally-supported, than low mass galaxies. This phenomenon, dubbed ``kinematic downsizing" (K12, S16), is present out to $z\,=\,2.5$. High mass galaxies reach a strong level of rotational support ($V_{rot}/S_{0.5}\,>\,1.3$) by a lookback time of $5-6$ Gyr, or $z\,\sim\,0.7$. Low mass galaxies reach the same degree of rotational support a few Gyrs later, by $z\,\sim\,0.2$. We fit a line to the median $V_{rot}/S_{0.5}$ using only the rising part of this relation, which covers $0.5\,<\,z\,<\,2.5$ in the high mass bin and $0.1\,<\,z\,<\,2.5$ in the low mass bin:

\begin{equation}\label{eq:4}
\frac{V_{rot}}{S_{0.5}} = a\left(\frac{t_{L}}{\textrm{Gyr}}\right)+b
\end{equation}
The best fit slope and intercept are $a=-0.074\,\pm\,0.010$ and $b=1.53\,\pm\,0.04$ for low mass galaxies and $a=-0.042\,\pm\,0.020$ and $b=1.59\,\pm\,0.08$ for high mass galaxies. The values for the best-fit slopes are similar and suggest that low mass and high mass galaxies develop rotational support at similar rates. We refit the high mass galaxies with a slope fixed to the best-fit slope of the low mass galaxies. At fixed slope, the high mass and low mass trends are offset by 4.3 Gyr. In other words, the assembly of rotational support in the high mass galaxies is, on average, followed $4-5$ Gyrs later by similar kinematic assembly in low mass galaxies.

As expected, we find inverted trends in the evolution of $\sigma_g/S_{0.5}$ (Figure \ref{fig:sig_v_S_z}, bottom right). On average, $\sigma_g/S_{0.5}$ declines with time for all galaxies and is always larger for low mass galaxies. Given that $\sigma_g$ itself does not depend on stellar mass (top right panel), its relative contribution to the total dynamical support will always be larger in galaxies with lower stellar mass. Beyond $z\,=\,$1.0, or a lookback time of $\sim\,9$ Gyr, low mass galaxies have large contributions from dispersion ($\sigma_g/S_{0.5}\,\gtrsim\,0.8$). They progressively decline in dispersion support with time up to the present day. At $z\,=\,0$, both low mass and high mass galaxies have low levels of dispersion support ($\sigma_g/S_{0.5} \le 0.4$) and high levels of rotational support ($V_{rot}/S_{0.5} \ge 1.3$). We fit a line to the median $\sigma_g/S_{0.5}$ over the declining part of the relation, covering $0.5\,<\,z\,<\,2.5$ for high mass galaxies and $0.1\,<\,z\,<\,2.5$ for low mass galaxies:

\begin{equation}\label{eq:5}
\frac{\sigma_g}{S_{0.5}} = a\left(\frac{t_{L}}{\textrm{Gyr}}\right)+b
\end{equation}
The best fit slope and intercept values are $a=0.063\,\pm\,0.01$ and $b=0.20\,\pm\,0.04$ for low mass galaxies and $a=0.056\,\pm\,0.02$ and $b=-0.02\,\pm\,0.11$ for high mass galaxies. As before, the slopes are statistically identical and suggest that the dispersion support declines at similar rates in both mass bins. We refit to the high mass galaxies with a slope fixed to low mass fit and measure an offset between the mass bins of 4.6 Gyrs. These results mirror our earlier conclusions and once again suggest that kinematic assembly in low mass galaxies is delayed from that in high mass galaxies by $\sim4-5$ Gyrs.

\section{The Fraction of Galaxies With Rotational Support}

We now investigate the fraction of star-forming galaxies with $V_{rot}/\sigma_g\,>\,1$ and $V_{rot}/\sigma_g\,>\,3$ as a function of time. These two values are arbitrary, but serve as relatively lenient benchmarks of rotational support. For perspective, sufficiently massive ($\log\,M_*/M_{\odot}\,>\,10$) star-forming disk galaxies in the local universe tend to have values of $V_{rot}/\sigma_g\,>\,5$ when measured through a hot ionized gas tracer such as H$\alpha$ \citep{2008MNRAS.390..466E} and $V_{rot}/\sigma_g\,>\,10$ when measured through a cold neutral gas tracer such as H\rom{1} \citep{2008AJ....136.2563W, 2008AJ....136.2648D}.

We divide our sample into 3 overlapping bins in fixed stellar mass:  $9\,<\,\log M_*/M_{\odot}\,<\,10$ (``low mass"), $9.5\,<\,\log M_*/M_{\odot}\,<\,10.5$ (``intermediate mass") and  $10\,<\,\log M_*/M_{\odot}\,<\,11$ (``high mass"). 

Figure \ref{fig:mass_settling} (left) shows the evolution of the fraction of galaxies with $V_{rot}/\sigma_g\,>\,1$. As first reported by K12 to $z\,=\,1.2$, this fraction increases with time for all galaxies and is always higher at higher stellar masses. The SIGMA sample extends these trends to $z\,=\,2$. The fraction of galaxies in SIGMA with $V_{rot}/\sigma_g\,>\,1$ for the low, intermediate and high mass bins is 49($\pm15$)$\%$, 51($\pm12$)$\%$, and 70($\pm13$)$\%$, respectively, at $z\,=\,2$. That is, about half of the star-forming galaxies in the low and intermediate mass bins have dominant contributions from $\sigma_g$. The standard error on the reported fractions are calculated via bootstrap resampling. Measurement uncertainties will push individual galaxies above or below the threshold in $V_{rot}/\sigma_g$ and we account for this by perturbing the values of $V_{rot}$ and $\sigma_g$ by their associated errors on each draw.

In Figure \ref{fig:mass_settling} (right), the fraction of galaxies which have $V_{rot}/\sigma_g\,>\,3$ is shown. The fraction of galaxies in SIGMA at $z = 2$ that meet this benchmark for the low, intermediate and high mass bins are only 15($\pm10$)$\%$, 11($\pm8$)$\%$, and 30($\pm13$)$\%$, respectively.

{ We fit the evolution of both of these fractions, $F$,  with:}

\begin{equation}\label{eq:6}
F\left(\frac{V_{rot}}{\sigma_g}\,>\,x\right) = a\left(\frac{t_{L}}{\textrm{Gyr}}\right)+b
\end{equation}

{ The best-fit slope and intercept for $F$($V_{rot}/\sigma_g\,>1$) are $a=-0.057\,\pm\,0.006$ and $b=1.08\,\pm\,0.04$ for low mass galaxies, $a=-0.056\,\pm\,0.009$ and $b=1.19\,\pm\,0.066$ for intermediate mass galaxies and $a=-0.045\,\pm\,0.009$ and $b=1.21\,\pm\,0.071$ for high mass galaxies. Similarly, the best-fit values for $F$($V_{rot}/\sigma_g\,>3$) are $a=-0.069\,\pm\,0.009$ and $b=0.84\,\pm\,0.07$ for low mass galaxies, $a=-0.085\,\pm\,0.007$ and $b=1.09\,\pm\,0.05$ for intermediate mass galaxies and $a=-0.08\,\pm\,0.01$ and $b=1.20\,\pm\,0.09$ for high mass galaxies.}

By $z=0.2$, more than $70\%$ of star-forming galaxies with $\log\,M_{*}/M_{\odot}> 9$ have reached $V_{rot}/\sigma_g\,>\,3$ and more than 90$\%$ have reached $V_{rot}/\sigma_g\,>\,1$. Even at $z=2$, most of the sample shows some degree of rotation. More than half of galaxies at this redshift have $V_{rot}/\sigma_g\,>\,1$. However, these galaxies are unlike local rotating galaxies, whose $V_{rot}/\sigma_g$ often exceeds 5.  { The redshift at which $35\%$ of galaxies have $V_{rot}/\sigma_g\,>\,3$ is $z=\,1$, 1.5 and 2.6 for the low, intermediate and high mass bins, respectively. }Kinematic analogs of local well-ordered disk galaxies are rare at $z\,=\,2$. At this redshift, only 7 (16\%) and 2 (4\%) of the galaxies in our sample reach $V_{rot}/\sigma_g\,>5$ and $\,>10$, respectively. 

\begin{figure}
\begin{centering}
\includegraphics[angle=0,scale=.58]{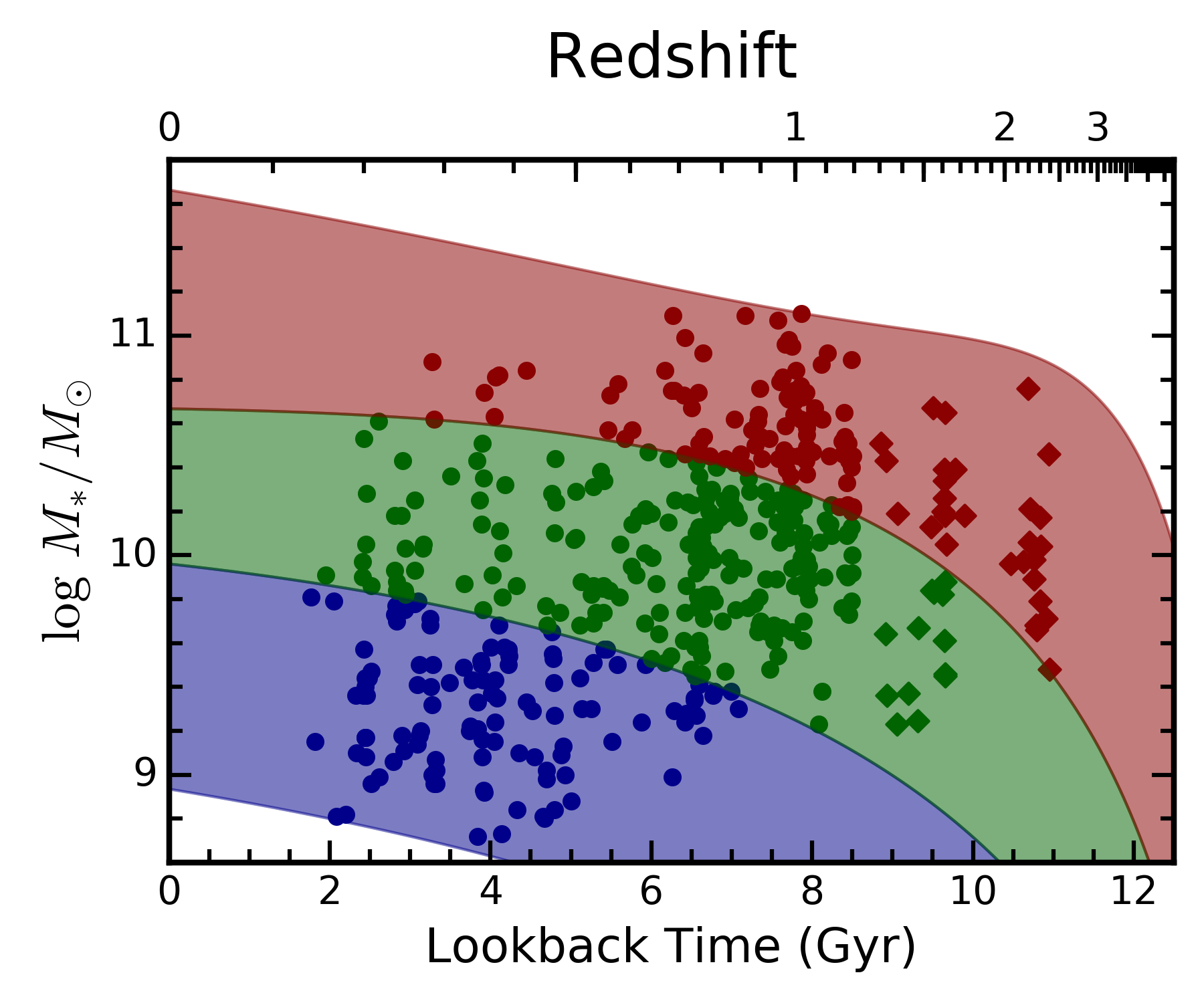}
\caption {Galaxies in our sample are linked in time using an abundance matching model from \citealt{2013MNRAS.428.3121M}. Three galaxy populations with $z\,=\,0$ halo masses of $11\,<\,M_{h}/M_{\odot}\,<\,11.5$ (blue), $11.5\,<\,M_{h}/M_{\odot} < 12.3$ (green) and $12.3\,<\,M_{h}/M_{\odot} < 14.5$ (red) are tracked in time. The average stellar mass evolution of the models is shown as a shaded swath.\vspace{0.1cm}}
\label{fig:ab_matching}
\end{centering}
\end{figure}

\begin{figure*}[ht]
\begin{centering}
\includegraphics[angle=0,scale=1.15]{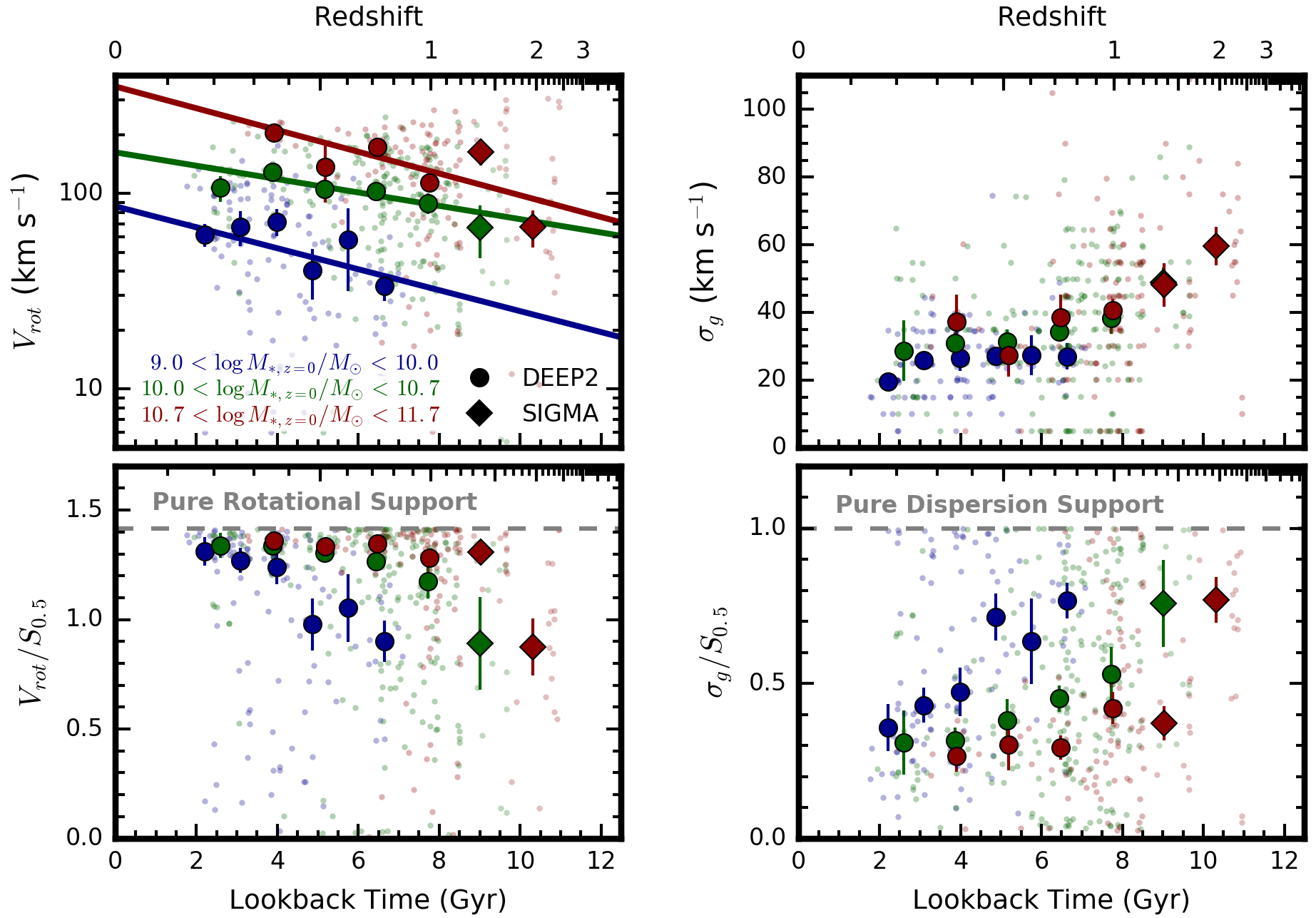}
\caption {The same as Figure \ref{fig:sig_v_S_z}, but for galaxy populations linked in time via abundance matching. All galaxy populations on average spin-up with time and decrease in disordered motions, i.e., increase in $V_{rot}$ and decline in $\sigma_g$. High mass galaxies reach strong levels of rotational support, i.e., $V_{rot}/S_{0.5}\,>\,1.3$ and $\sigma_g/S_{0.5}\,<\,0.4$, by $z\,=\,1.5$ on average. Low and intermediate mass galaxies reach similar levels of rotational support several Gyrs later. The color scheme is the same as in Figure \ref{fig:ab_matching} with dark blue, green and red representing the low ($\log M_{*, z = 0}/M_{\odot}\sim9.4$), intermediate ($\log M_{*, z = 0}/M_{\odot}\sim10.3$) and high ($\log M_{*, z = 0}/M_{\odot}\sim11.1$) mass abundance matched populations, respectively. The small points are individual galaxies and large points are the medians in bins of lookback time. The lines are the best-fit relations to the median points of each mass bin. \vspace{0.1cm}}
\label{fig:Vsig_time}
\end{centering}
\end{figure*}

\begin{figure}[ht]
\begin{centering}
\includegraphics[angle=0,scale=1.1]{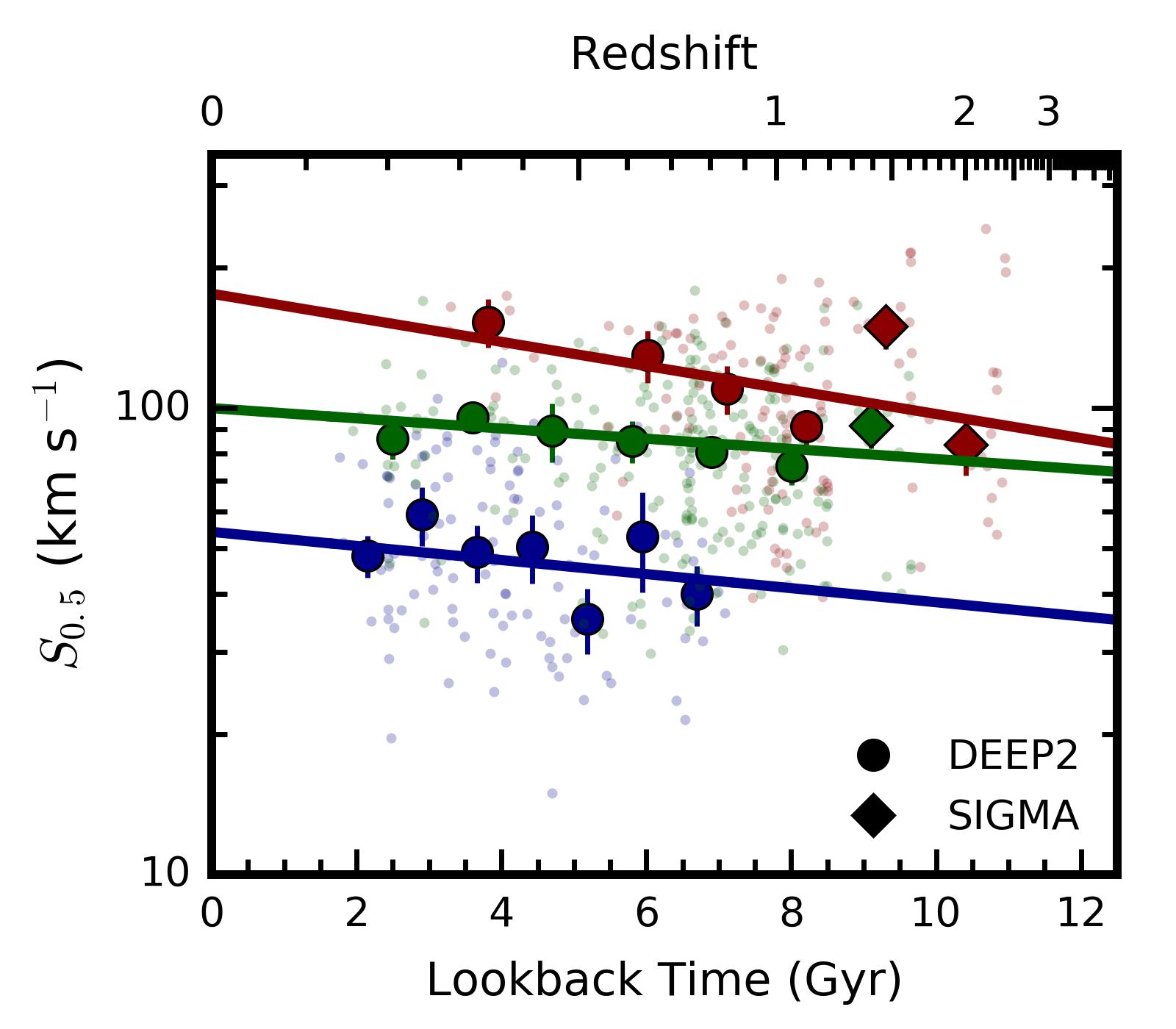}
\caption {The same as Figure \ref{fig:S_time}, but for galaxy populations linked in time via abundance matching. The quantity $S_{0.5}$ increases with time for all galaxy populations, indicating that their potential wells grow with time. The lines, points and color scheme are the same as in Figure \ref{fig:Vsig_time}. \vspace{0.1cm}}
\label{fig:S_abmatch_time}
\end{centering}
\end{figure}

\subsection{Comparison With the Literature}

In Figure \ref{fig:mass_settling} (left), we compare our measurements with those in the literature. Any comparison between surveys is complicated due in large part to differences in sample selection and measurement techniques. We refrain from performing detailed corrections between the samples and only report here a straightforward comparison. We include measurements from the following surveys: KROSS at $z\,\sim1$ \citep{2016MNRAS.457.1888S}, KMOS-3D at $z\,\sim1$ and $z\,\sim2$ \citep{2015ApJ...799..209W}, MASSIV at $z\,\sim1.3$ \citep{2012A&A...539A..91C}, AMAZE/LSD at $z\,\sim3$ \citep{2011A&A...528A..88G}, { and the KDS survey at $z\,\sim\,3.5$ \citep{2017arXiv170406263T}.} The adopted measurements are briefly described below. To account for differences in the stellar mass of galaxies in each sample, we color-code each by its approximate mass coverage. Our sample and the literature measurements are in good agreement once we account for stellar mass.

The KROSS sample covers the redshift range $0.6\,<\,z\,<\,1.0$ and a stellar mass range of $9.0\,\lesssim\,\log\,M_{*}/M_{\odot}\,\lesssim\,11.0$. A catalog of kinematic measurements for 586 typical star-forming galaxies was released in \citet{2017MNRAS.467.1965H}. We use this catalog to calculate the fraction of resolved galaxies in KROSS which have rotation velocities that exceed their intrinsic velocity dispersions. We adopt mass bins which overlap our sample and these are shown as separate points in Figure \ref{fig:mass_settling}. The first-year sample from the KMOS-3D survey contains 191 star-forming galaxies in two redshift intervals: $0.7\,<\,z\,<\,1.1$ and $1.9\,<\,z\,<\,2.7$. These galaxies cover a stellar mass range of $9.6\,\lesssim\,\log\,M_{*}/M_{\odot}\,\lesssim\,11.1$ at $z\,\sim1$ and $10\,\lesssim\,\log\,M_{*}/M_{\odot}\,\lesssim\,11.2$ at $z\,\sim2$. \citet{2015ApJ...799..209W} report the fraction of galaxies in their sample with rotationally-supported kinematics as 93$\%$ and 74$\%$ at $z\,\sim1$ and $z\,\sim2$, respectively. We include the sample of 48 MASSIV galaxies from \citet{2012A&A...546A.118V}, covering $1\,<\,z\,<\,1.6$ and a stellar mass range of $9.5\,\lesssim\,\log\,M_{*}/M_{\odot}\,\lesssim\,11$. We adopt the $V/\sigma$ measurements in their Table 2 and calculate a rotation-dominated fraction of 65$\%$. We include the high redshift sample from the AMAZE/LSD survey of 33 galaxies over $2.6\,<z<\,4.9$ and a stellar mass range of $9.0\,\lesssim\,\log\,M_{*}/M_{\odot}\,\lesssim\,11.0$. \citet{2011A&A...528A..88G} report that 33$\%$ of the AMAZE galaxies can be classified as rotationally-supported. { Finally, we include results from the KMOS Deep survey (KDS) at $z\,\sim\,3.5$ \citep{2017arXiv170406263T} which covers a mass range of $8\,<\,\log M_*/M_{\odot}\,<\,11$. \citealt{2017arXiv170406263T} report that 13/32 of the isolated field galaxies in this sample are rotation-dominated. We remove the 4 galaxies with $\log M_*/M_{\odot}\,<9$, 3 of which are dispersion-dominated, and compute a rotation-dominated fraction of 43$\%$ over our mass range.}

\section{Tracing Galaxy Populations with Abundance Matching}

In previous sections of this paper, galaxies trends are reported for a fixed stellar mass. However, as a natural consequence of ongoing star formation and mass accretion, the galaxies in our sample will grow in stellar mass with time and migrate between mass bins. As such, the trends discussed so far should not be interpreted as tracks for individual galaxy populations. In order to trace the evolution of a galaxy population from $z=2.5$ to $z=0$, we adopt a model linking high redshift galaxies with their more massive low redshift descendants. 

Galaxy populations are tracked in time using the multi-epoch abundance matching (MEAM) model \citep{2013MNRAS.428.3121M}, following \citet{2015ApJ...803...26P}. { The conclusions of this section are similar if we adopt other abundance matching models from the literature (e.g., \citealt{2017arXiv170304542R}).} The abundance matching technique assumes that there is a monotonic relation between the stellar mass of a galaxy and its halo mass, { with the most massive galaxies residing in the most massive halos.} Observed stellar mass functions are then rank assigned to simulated halo mass functions at all redshifts and galaxy populations are tracked in time.

\citet{2013MNRAS.428.3121M} adopt halo and subhalo evolution from the Millenium simulation and stellar mass functions up to $z\,\sim\,4$ from several surveys in the literature \citep{2008ApJ...675..234P, 2009MNRAS.398.2177L, 2012A&A...538A..33S}. Fitting functions are provided for the mass accretion history, star-formation history and stellar mass loss rates of galaxies as a function of mass. We integrate these prescriptions in time to track the average stellar mass evolution of galaxy populations. We divide our observational sample into 3 populations of $z\,=\,0$ halo masses of $11\,<\,M_{h}/M_{\odot}\,<\,11.5$, $11.5\,<\,M_{h}/M_{\odot} < 12.3$ and $12.3\,<\,M_{h}/M_{\odot} < 14.5$. These correspond to typical $z\,=\,0$ stellar masses of $\log M_{*}/M_{\odot}\sim$ 9.4, 10.3, and 11.1, respectively.

The average stellar mass evolution for these three populations is shown in Figure \ref{fig:ab_matching}. The SIGMA survey does not cover $\log M_*/M_{\odot}\,<\,9$ and so the lowest mass bin does not have coverage beyond $z\,\sim\,1$. Similarly, the intermediate mass bin only extends to $z\,\sim\,1.5$. We use these tracks to link galaxies in our sample with their appropriate descendants.

\subsection{Increasing $V_{rot}$, Decreasing $\sigma_g$, and Increasing $S_{0.5}$ With Time}
Using the abundance matching technique to link galaxy populations in time, we now revisit the evolution of $V_{rot}$, $\sigma_g$ and $S_{0.5}$ in Figures \ref{fig:Vsig_time} and \ref{fig:S_abmatch_time}.

The most striking difference is in the evolution of $V_{rot}$. At fixed stellar mass we measured mild to no evolution in $V_{rot}$. However, in Figure \ref{fig:Vsig_time} (top left) we show that $V_{rot}$ strongly evolves for the evolving galaxy population: \emph{star-forming galaxies spin-up with time on average}. We use Eq. \ref{eq:1} and fit to the median evolution of $V_{rot}$ in each mass bin. The best-fit slope and intercept values are $a\,=-0.047\,\pm0.012$ and $b\,=0.90\,\pm0.054$ for low mass galaxies, $a\,=-0.053\,\pm0.010$ and $b\,=1.11\,\pm0.04$ for intermediate mass galaxies, and $a\,=-0.058\,\pm0.090$ and $b\,=1.27\,\pm0.070$ for high mass galaxies. The best-fit slopes are similar and indicate that star-forming galaxies in this mass range have a typical $V_{rot}$ doubling timescale of $\sim$6 Gyrs, similar to the half-life timescale of $\sigma_g$ at fixed stellar mass (\S3.2). 

At fixed galaxy population, we again find that $\sigma_g$ dramatically declines with time (Figure \ref{fig:Vsig_time}, top right). Intermediate mass and high mass galaxies trace each other back to $z\,=\,1.5$, rising from 30 km s$^{-1}$ at $z\,=\,0.2$ to 45 km s$^{-1}$ at $z\,=\,1.5$. The high mass galaxies, for which we have measurements out to high redshift, reach typical values of $\sigma_g\,=\,60$ km s$^{-1}$ at $z\,=\,2$.

The evolution of S$_{0.5}$ for the abundance matched populations is shown in Figure \ref{fig:S_abmatch_time}. At fixed stellar mass, we found a mild decline in $S_{0.5}$ with time. At fixed galaxy population, we find that $S_{0.5}$ increases with time for all galaxies. This result suggests that, perhaps unsurprisingly, galaxy potential wells steepen as they acquire mass. The quantity $\sigma_g$ declines with time on average and so the additional dynamical support comes from the increase in $V_{rot}$. 

As before, we use Eq. \ref{eq:3} and fit to the median $S_{0.5}$ evolution for each mass bin. The best-fit slope and intercept values are $a\,=-0.015\,\pm\,0.019$ and $b\,=1.73\,\pm\,0.08$ for the fixed population at low mass, $a\,=-0.01\,\pm\,0.01$ and $b\,=2.0\,\pm\,0.05$ at intermediate mass, and $a\,=-0.03\,\pm\,0.04$ and $b\,=2.24\,\pm\,0.28$ at high mass.

We normalize $V_{rot}$ and $\sigma_g$ by $S_{0.5}$ (Figure \ref{fig:Vsig_time}, bottom panels) and find the same general behavior as before: while all galaxy populations increase in rotational support with time (rising in $V_{rot}/S_{0.5}$ and declining in $\sigma_g/S_{0.5}$), massive galaxies are the most rotationally supported at all times on average. The high mass population reaches a strong level of rotational support ($V_{rot}/S_{0.5}\,>\,1.3$) as far back as $z\,=\,1.5$, while the intermediate mass and low mass populations reach the same degree of rotational support at $z\,=\,0.5$ and $z\,=\,0.2$, respectively.

\begin{figure*}[ht]
\begin{centering}
\includegraphics[angle=0,scale=0.42]{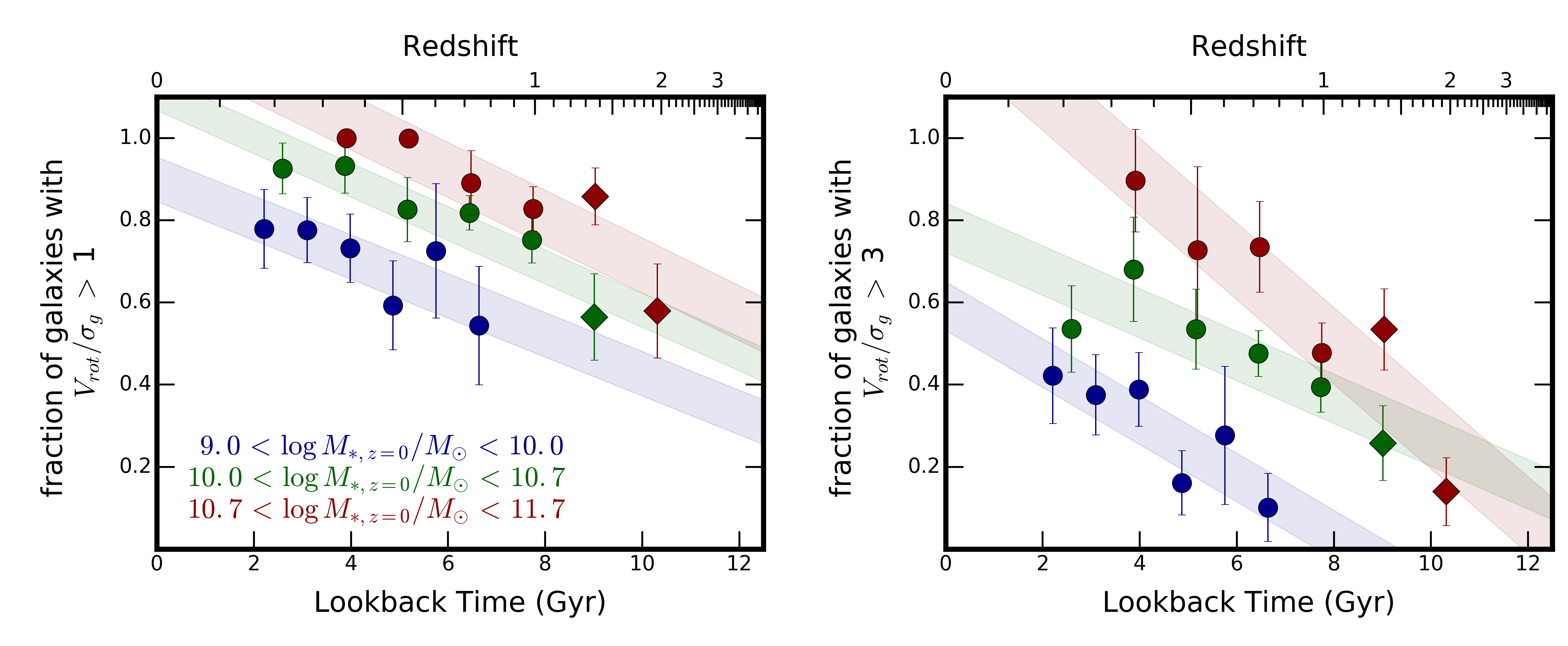}
\caption {The same as Figure \ref{fig:mass_settling}, but for galaxy populations linked in time via abundance matching. For all three galaxy populations (low, intermediate and high mass), the fractions of galaxies with $V_{rot}/\sigma_g\,>1$ and 3 increase significantly with time. At any point in time, higher mass populations have a higher fraction of galaxies meeting these criteria. The fraction of galaxies with $V_{rot}/\sigma_g\,>\,3$ drops below 30$\%$ at $z\sim0.4$ for the low mass galaxy population (blue) and at $z\sim1.5$ for the more massive galaxy populations (green and red). As in Figure \ref{fig:mass_settling}, the solid points show measurements from DEEP2 (circles) and SIGMA (diamonds), and the error bars are from bootstrap resampling. The shaded swaths represent the uncertainty on the intercept in a linear fit to the points. \vspace{0.1cm}}
\label{fig:abmatch_settling}
\end{centering}
\end{figure*}

\subsection{Fraction of galaxies with $V_{rot}/\sigma_g\,>\,1$ and 3 with time}

{ In Figure \ref{fig:abmatch_settling}, we use the linked galaxy populations to revisit how the fraction of galaxies with $V_{rot}/\sigma_g\,>\,1$ and 3 evolves with time and galaxy mass. 

As before, at a fixed stellar mass, we find trends that are smooth in both mass and redshift. These results are consistent with kinematic downsizing: more galaxies tend to become rotation-dominated with time and, at all times, galaxies are more likely to be rotation-dominated at higher mass. As before, this result is due in part to the large decline in $\sigma_g$ with time and its weak dependance on mass. Additionally, unlike the case at fixed stellar mass, there is a contribution from the steep rise in $V_{rot}$ with time in all galaxy populations and its strong dependance on mass.

We use Eq. \ref{eq:6} to fit to the evolution in each mass bin. The best-fit slope and intercept for $F$($V_{rot}/\sigma_g\,>1$) are $a=-0.047\,\pm\,0.012$ and $b=0.90\,\pm\,0.05$ at low mass, $a=-0.053\,\pm\,0.007$ and $b=1.11\,\pm\,0.04$ at intermediate mass and $a=-0.06\,\pm\,0.009$ and $b=1.27\,\pm\,0.07$ at high mass. For $F$($V_{rot}/\sigma_g\,>3$) the best-fit values are $a=-0.07\,\pm\,0.013$ and $b=0.59\,\pm\,0.06$ at low mass, $a=-0.052\,\pm\,0.01$ and $b=0.78\,\pm\,0.06$ at intermediate mass and $a=-0.10\,\pm\,0.01$ and $b=1.32\,\pm\,0.09$ at high mass.

The majority of galaxies in these populations have some degree of rotational support (Figure \ref{fig:abmatch_settling}, left). More than half of the galaxies in the low mass bin have $V_{rot}/\sigma_g\,>\,1$ at $z\,=\,1$ and the intermediate and high mass bins reach similar values at $z\,=\,2$. 

However, it is much less common for galaxies to reach $V_{rot}/\sigma_g\,>\,3$ at high redshifts (Figure \ref{fig:abmatch_settling}, right). Less than $30\%$ of galaxies in the low mass population have $V_{rot}/\sigma_g\,>\,3$ beyond $z\,\sim\,0.4$ (Figure \ref{fig:abmatch_settling}, right), indicating that it would have been exceedingly rare for today's LMC-mass galaxies to have had strong rotational support at redshifts greater than 1. The intermediate and high mass populations drop below 30$\%$ around $z\,\sim\,1.5$. The most massive star-forming disks in the local universe ($\log {M_{*,z=0}/M_\odot}\,>\,10.0$; i.e., Milky-Way and M31-mass galaxies) were likely weakly rotationally-supported, $V_{rot}/\sigma_g\,<\,3$, at the peak of cosmic star-formation.}

\section{Discussion and Conclusions}

To explore the evolution of internal galaxy kinematics over a continuous period of time from $z\,=\,2.5$ to today, measurements from the DEEP2 survey ($0.1\,<\,z\,<\,1.2$; \citealt{2007ApJ...660L..35K, 2012ApJ...758..106K}) are combined with those from the SIGMA survey ($1.3\,<\,z\,<\,2.5$; \citealt{2016ApJ...830...14S}). The full sample contains 507 star-forming galaxies and covers a mass range of $9\,<\,\log M_{*}/M_{\odot}\,<\,11$. Our sample is homogenous and representative of galaxies on the star-formation main sequence. There is no selection on morphology. Kinematics were measured using the same software in both surveys. This data set is the largest contiguous and morphologically unbiased sample of kinematics in the literature and it extends the work of \citet{2012ApJ...758..106K} from $z\,=\,1.2$ into the peak of cosmic star-formation at $z\,=\,2$.

Kinematics are measured from rest-optical emission lines (H$\alpha$, \OIII$\lambda$ 5007) in slit spectra, which trace hot ionized $10^4$ K gas. Two kinematic quantities are measured: the rotation velocity $V_{rot}$, which traces ordered motions, and the gas velocity dispersion $\sigma_g$, which traces disordered motions. Both of these quantities provide dynamical support, and when combined into the quantity $S_{0.5}\,=\,\sqrt{0.5V_{rot}^2+\sigma_g^2}$, trace the depth of galaxy potential wells \citep{2006ApJ...653.1027W, 2007ApJ...660L..35K, 2010ApJ...710..279C}.

Average trends in $\sigma_g$ and $V_{rot}$ with time are presented for galaxies at fixed stellar mass. We find a smooth and dramatic decline in $\sigma_g$ since z = 2.5. The time for $\sigma_g$ to decline by a factor of two, or the half-life time, is $\sim$ 6 Gyrs. Remarkably, this evolution is similar for all masses. Over the same time period, $V_{rot}$ increases in galaxies with low stellar mass ($9\,<\,\log M_*/M_{\odot}\,<\,10$), but does not evolve in galaxies with high stellar mass ($10\,<\,\log M_*/M_{\odot}\,<\,11$). 

The fractions of dynamical support provided by rotation and dispersion are measured by taking the ratios of $V_{rot}$ and $\sigma_g$ to $S_{0.5}$.  At a fixed stellar mass, all galaxies on average rise in $V_{rot}/S_{0.5}$ with time, i.e. they become increasingly supported by rotation. In the same vein, all galaxies on average decline in $\sigma_g/S_{0.5}$ with time, i.e., they become decreasingly supported by dispersion. While the slopes of these trends ($V_{rot}/S_{0.5}$ and $\sigma_g/S_{0.5}$ with time) are similar for all masses, high mass galaxies have higher $V_{rot}/S_{0.5}$ and lower $\sigma_g/S_{0.5}$ at all times. This indicates a time delay in the development of rotational support between low mass and high mass galaxies, with low mass trailing high mass by a few Gyrs (i.e., ``kinematic downsizing"; \citealt{2012ApJ...758..106K, 2016ApJ...830...14S}). This result arises from the independence of $\sigma_g$ on mass: a given value of $\sigma_g$ assumes a more dominant role in the shallower potential wells that host low mass galaxies.

These results are complimentary to studies which track morphological regularity, tracing the distribution of stars in galaxies, as a function of mass and redshift. Morphologically regular stellar disks are increasingly common at low redshift ($z\le1.5$) and high mass (e.g., \citealt{2013MNRAS.433.1185M, 2014ApJ...792L...6V, 2016MNRAS.462.4495H}).  For example, \citet{2016MNRAS.462.4495H} find that as much as 80$\%$ of the stellar mass density at $z\,=\,2$ resides in galaxies with disturbed/irregular morphologies.

Galaxies grow in stellar mass with time and so these trends, which are at fixed stellar mass, should not be interpreted as tracks for individual galaxy populations. To link galaxy populations in time, we adopt an abundance matching model. Doing so, we find that $V_{rot}$ increases, $\sigma_g$ decreases, and the potential well depth $S_{0.5}$ increases with time for all galaxies. The simultaneous rise in $S_{0.5}$ and $V_{rot}$ indicates that galaxies spin-up as they assemble their mass. On average, a star-forming galaxy of Milky-Way mass which is today strongly supported by rotation ($\sigma_g/S_{0.5}\,<\,0.3$) was likely strongly supported by dispersion at $z\,=\,2$ ($\sigma_g/S_{0.5}\,=\,0.7$), at a time when it was also much less massive. These results suggest that the assembly of stellar mass and the assembly of rotational support were contemporaneous and that the processes which govern stellar mass assembly at late times should also promote disk formation. { These processes likely include, but are not limited to, the accretion of mass and (re-)formation of disks in gas-rich interactions (e.g., \citealt{2006ApJ...645..986R, 2009MNRAS.398..312G, 2015MNRAS.451.4290S, 2017MNRAS.467.3083R}), the cessation of misaligned or destructive mergers (e.g., \citealt{2013MNRAS.434.3142A}), and a transition in the mode of star-formation from one which is bursty and violent at high redshift to one which is relatively calm and stable at low redshift (e.g., \citealt{2017MNRAS.467.2664C, 2017MNRAS.467.2430M}).}

At $z\,=\,0.2$, the vast majority ($90\%$) of star-forming galaxies are rotation-dominated ($V_{rot}>\sigma_g$). By $z\,=\,2$, the percentage of galaxies with $V_{rot}>\sigma_g$ has declined to 50$\%$ for low mass systems ($10^{9}-10^{10}\,M_{\odot}$) and 70$\%$ for high mass systems ($10^{10}-10^{11}M_{\odot}$). These measurements are consistent those in the literature once stellar mass is taken into account. We consider a stronger criterion for rotational support, $V_{rot}\,>\,3\,\sigma_g$, and find that the fractions of galaxies meeting this threshold drop below 35$\%$ for all masses. For perspective, most sufficiently massive galaxies in the local universe ($\log M_*/M_{\odot}\,>\,10$) have $V_{rot}/\sigma_g\,>\,5$ while the most massive disks regularly have $V_{rot}/\sigma_g\,>\,10$.

In conclusion, strong rotational support was exceedingly rare at $z\,=\,2$ and the kinematic characteristics of local disks, i.e., low $\sigma_g$ and high $V_{rot}/\sigma_g$, were only just beginning to emerge. This epoch is one of disk assembly, as star-forming galaxies are rapidly assembling stellar mass and beginning to develop the first well-ordered disks.

\section*{Acknowledgements}
The data presented herein were obtained at the W.M. Keck Observatory, which is operated as a scientific partnership among the California Institute of Technology, the University of California and the National Aeronautics and Space Administration. We wish to extend thanks to those of Hawaiian ancestry on whose sacred mountain we are privileged guests.  We thank the referee for providing a useful report which has improved this paper. RCS and SAK would like to acknowledge an RSAC grant from STScI. GFS appreciates support from a Giacconi Fellowship at the Space Telescope Science Institute, which is operated by the Association of Universities for Research in Astronomy, Inc., under NASA contract NAS 5-26555. SF and DK acknowledge support from NSF grant AST-0808133 to UCSC. This work has made use of the Rainbow Cosmological Surveys Database, which is operated by the Universidad Complutense de Madrid (UCM), partnered with the University of California Observatories at Santa Cruz (UCO/Lick,UCSC). This research made use of Astropy, a community-developed core Python package for Astronomy \citep{2013A&A...558A..33A}.

\end{document}